\documentclass[a4paper,12pt,final]{article}

\usepackage[notcite,notref]{showkeys}
\usepackage{amsfonts,amssymb,amsthm}
\usepackage{amsmath,amssymb}
\usepackage{hyperref}
\usepackage{bold-extra}
\usepackage{subfig}
\usepackage{tikz}
\usepackage{tikz-3dplot}
\usepackage{pgfplots}
\usetikzlibrary{decorations.pathmorphing} 
\usepackage[font={small,rm}]{caption}
\usetikzlibrary{calc}

\newcommand{\COMMENT}[1]{}

\usepackage{amssymb,amsmath,mathrsfs,amsthm,color,srcltx,fullpage,pdfpages,bm}
\usepackage{graphicx}
\usepackage{hyperref}

\newcommand\bs[1]{\boldsymbol {#1}}

\newcommand\be{\boldsymbol e}

\newcommand\bR{\boldsymbol R}

\newcommand\bx{\bs x}
\newcommand\br{\bs r}
\newcommand\ba{\bs a}

\newcommand {\bydef}{\,\raise.07485ex\hbox{:}\kern-.025em\hbox{=}\,}
\newcommand{\vth}{\vartheta}          

\newcommand {\lamp} {\lambda_\|}
\newcommand {\lamo} {\lambda_\bot}

\author{Giuseppe Tomassetti\footnote{DICII Department, University of Rome Tor Vergata. Via Politecnico 1, 00133 Rome, Italy. Email: \texttt{tomassetti@ing.uniroma2.it}} \, and Valerio Varano\footnote{LaMS Modelling and Simulation Lab, University of Rome Roma Tre. Via della Madonna dei Monti,40, 00184 Rome, Italy. Email: \texttt{valerio.varano@uniroma3.it}}}
\title{Capturing the helical to spiral transitions in thin ribbons of nematic elastomers.}

\begin{document}
\date{July 29, 2016}
\maketitle
\begin{abstract}
We provide a quantitative description of the helicoid--to--spiral transition in thin ribbons of nematic elastomers using an elementary calculation based on a Koiter-type plate with incompatible reference configuration.  Our calculation confirms that such transition is ruled by the competition between stretching energy and bending energy.
\end{abstract}


\tableofcontents

\section{Introduction}
During the last decade engineers, physicists, and applied mathematicians have been devoting significant research effort to programming and controlling the shape change of devices made of soft materials, that is, materials that can be easily deformed by the application of external stimuli. The ability of these devices to attain extreme shape changes in a controlled or programmed fashion make them ideal candidates for smart actuators, nanorobots, devices for drug delivery, etc. \cite{FuscoHPPHUSPSPo2015Aami}. Moreover, thin structures that can fold in a programmed and controlled way provides an attractive opportunity for the microfabrication of three-dimensional objects from flat thin films \cite{Ionov2013PR}.

Controlling shape change is considerably easier when the device is a thin or slender object, such as a plate or a rod. The reason is two-fold: first, thin structures display a significant compliance and the strains required to attain large displacements are small if compared with bulky objects --- this why the most dramatic deformations occurring in nature are observed in thin objects; second, the pointwise value of the strain field is easier to control for thin objects, rather than for bulky objects. For these reasons, the combination of slenderness and softness is  the key feature towards the realization of devices that can undergo extreme changes of geometry and shape with minimal input of energy \cite{AmarG2005Growth,GorielyBenAmar2005,MoultLG2013JMPS,TieroTmorphoelastic}. 

Elastomeric materials appear to be ideal candidates to realize soft devices, thanks to their ability of undergoing substantial strains, their biocompatiobility, and their ease of manufacturing. For these materials, shape change can be triggered through a variety of external stimuli and can be programmed by a suitable spatial modulation of their physical properties. One method to enforce shape changes in an elastomeric thin device is to induce a non-uniform swelling across its thickness by a careful control of its chemical environment \cite{DiasHS2011PRE,LucanN2012IJSS,lucantonio2013transient,WuMGTNSK2013Nc}. Shape control through temperature change can instead be realized using nematic elastomers. Above its critical temperature, a nematic elastomer displays isotropic behavior. However, when the temperature goes below its critical value, the isotropic--to--nematic phase transformation induces spontaneous strain whose principal directions depend on the orientation of the nematic director (see Fig. \ref{Fig_1} in Sec. \ref{sec:2} below). By imprinting in the material a non-constant nematic director field, a non-uniform strain can be induced, which can result in a variety of shape changes. 

The experiments in \cite{Sawa2010,sawa2011shape,sawa2013shape} describe shape change in thin strips of nematic elastomers with various orientation pattern. Particularly interesting is the twisted pattern devised in  \cite{sawa2011shape}, which features a chiral arrangement of the nematic director (see Fig. \ref{fig1} in Sect. \ref{sec:3} below). When imprinted with such pattern the strip can attain two different configurations: within a certain range of ambient temperature, the specimen attains a helicoidal shape (see Fig. \ref{fig:dummy}.b in Sect. \ref{sec:3}); outside that range, the specimen takes a spiral-like shape (Fig. \ref{fig:dummy}.c).

The temperature range where the strip remains in the helicoid configuration has a specific dependence on the ratio $h/w$ between its thickness $h$ and its width $w$.  A one-dimensional model providing theoretical explanation of such dependence was provided in \cite{sawa2011shape}. A numerical validation using three-dimensional elasticity with distortions was given in \cite{TeresV2013SM}. In this paper we provide a quantitative description of the helicoid--to--spiral transition using an elementary calculation based on a Koiter-type plate with distortions. The phase diagram we obtain from our calculation is in fairly good agreement with that in \cite{sawa2011shape}. Our result confirms that the switching from one configuration to another is the result of the competition between stretching energy and bending energy. A possible developments of this approach might be a further dimension reduction where the width of the strip tends to null, using a variational approach as in \cite{FreddHMP2016apa,AgostDK2016Shape} or a deductive approach based on reduced kinematics in the spirit of \cite{gabriele20141d,gabriele20161d}.  

The paper is organized as follows. In Section 2 we provide physical background on nematic elastomers and on the three--dimensional theory developed in  \cite{Teresi2009}. In Section 3 we specialize the theory to thin strips of nematic elastomers with twisted director pattern. In Section 4 we compute the energy associated to the helicoid and to spiral configuration and we compute a phase diagram which shows what particular configuration arises as function of the strip's geometry. The shell theory we use to perform our calculation is derived  in the Appendix through a formal dimension-reduction argument procedure to the three dimensional model proposed in \cite{Teresi2009}.

\section{Physical background}\label{sec:2}
A distinctive feature of elastomers is that they can undergo substantial reversible strain. For example, rubber specimens can be stretched by several times their length without breaking and they recover their original shape when unloaded. These features are explained by a closer inspection of the microscopic structure of these materials, which is substantially different from that of a crystalline solids. 

The elementary units forming an elastomer are not arranged in a lattice, but linked in pairs to form long polymer chains. In the absence of external forces, these chains coil up so that their end-to-end separation is several times smaller than their length. What confers elastomers their solid-like behavior is the presence of several cross links between polymeric chains, usually obtained by modifying the polymerization process through the introduction, in addition to the usual bi-valent monomeric segments, of a certain amount of trivalent segments, or by adding a cross-linking agent to a system of already-assembled polymeric chains. In its stress-free configuration, the polymeric network forming an elastomeric material does not have any directional ordering. From the mechanical standpoint it may be described as a Neo-Hookean material, characterized by an elastic per unit reference volume of the form:
\begin{align}\label{eq:104}
& \widehat\sigma(\mathbf C)=\frac G 2 
\big(\mathbf C\cdot\mathbf I-\log({\rm det}\mathbf C)-3\big)+\frac k 8({\rm det}\mathbf C-1)^2,
\end{align}
where  $\mathbf C$  is the Cauchy--Green strain.

Since the bulk modulus $k+2/3G$ is three orders of magnitude larger than the shear modulus $G$, elastomers are nearly incompressible. However, if a dry elastomer is placed in a solvent, it attains chemical equilibrium by adsorbing the solvent and by increasing its volume. The process is entirely reversible: when dried, the elastomer will recover its original shape.

The driving force behind the swelling is the entropy of mixing \cite{Flory1942TJocp}, which facilitates the dispersion of the polymer molecules in the solvent: in fact, if the polymer molecules making up the elastomer were not joined together by cross links, they would dissolve in the solvent, occupying all the available space. It is exactly the competition between elasticity and entropy of mixing that determines the dispersion of network in the solvent or, equivalently, the amount of solvent that permeates the elastomeric network. 

\begin{figure}[h]
\begin{center}
\begin{minipage}{0.8\linewidth}
\begin{center}
\begin{tikzpicture}[scale=1.725]
   \shade[left color=yellow,right color=blue] (0,0) ellipse (0.7 and 0.7);
   %
   \shade[ball color=red, rotate=30] (0.,0.) ellipse (0.15 and 0.05);
   \shade[ball color=red, rotate=45] (0.1,0.4) ellipse (0.15 and 0.05);
   \shade[ball color=red, rotate=120] (0.1,-0.5) ellipse (0.15 and 0.05);
   \shade[ball color=red, rotate=50] (-0.2,-0.20) ellipse (0.15 and 0.05);
   \shade[ball color=red, rotate=-50] (-0.3,-0.30) ellipse (0.15 and 0.05);
   \shade[ball color=red, rotate=-30] (0.5,-0.20) ellipse (0.15 and 0.05);
   \shade[ball color=red, rotate=75] (-0.0,-0.50) ellipse (0.15 and 0.05);
   \shade[ball color=red, rotate=95] (-0.35,0.30) ellipse (0.15 and 0.05);
   \shade (0.3,0.40) [rotate=20] ellipse (0.15 and 0.05) [ball color=red] ;
   \draw [<->] (-0.7,-0.8) -- (0,-0.8) node[below] {$\ell_o$}-- (0.7,-0.8);
   \draw [<->] (1.9,-0.8) -- (2.9,-0.8) node[below] {$\ell_o\,\lamp$} --(3.7,-0.8);
   \draw [<->] (1.6,-0.6) -- (1.6,0) node[above,rotate=90] {$\ell_o\,\lamo$} --(1.6,0.6);
   %
   %
   \shade[left color=yellow, right color=blue](2.8,0) ellipse (1. and 0.6);
   %
   \shade[ball color=red] (2.8,0.) ellipse (0.15 and 0.05);
   \shade[ball color=red, rotate=15] (3.4,-0.80) ellipse (0.15 and 0.05);
   \shade[ball color=red, rotate=-15] (3.1,0.75) ellipse (0.15 and 0.05);
   \shade[ball color=red, rotate=-25] (2.5,1.65) ellipse (0.15 and 0.05);
   \shade[ball color=red, rotate=-30] (2.3,1.) ellipse (0.15 and 0.05);
   \shade[ball color=red, rotate=15] (2.,-.65) ellipse (0.15 and 0.05);
   \shade[ball color=red, rotate=-15] (2.2,.85) ellipse (0.15 and 0.05);
   \shade[ball color=red, rotate=20] (2.5,-.55) ellipse (0.15 and 0.05);
   \shade[ball color=red, rotate=20] (2.7,-1.5) ellipse (0.15 and 0.05);
\end{tikzpicture}
\end{center}

\caption{\label{Fig_1} Left: the stress-free shape of a homogeneous spherical sample of diameter $\ell_0$ in the isotropic phase, above the critical temperature $T_{\sf NI}$. Right: the stress-free shape of the same sample when its temperature drops below $T_{\sf NI}$ is a prolate spheroid whose polar axis is aligned with the prevailing mesogen direction.} 
\end{minipage}
\end{center}
\end{figure}
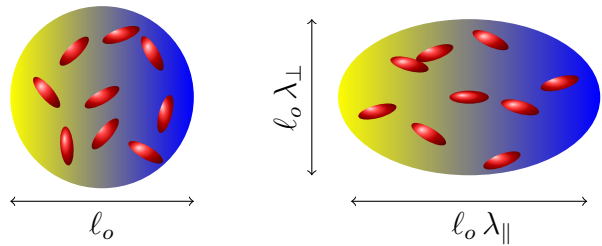

\medskip
\emph{Nematic elastomers} are networks of polymer chains obtained either by assembling mesogenic monomers, or by attaching mesogens to a conventional polymeric backbone. Mesogens have an elongated shape and, in the so--called \emph{nematic phase}, exhibit long-range directional order, their preferred direction at a given material point being identified by a \emph{tensorial order parameter}:
\begin{equation}\label{eq:88}
  \mathbf N(\bm x)=\bm d(\bm x)\otimes\bm d(\bm x),
\end{equation}
where $\bm d$ is a unit vector that can possibly depend on the referential position $\bm x$ of the material point. Nematic ordering is entropy driven: alignment between molecules, which by itself would decrease entropy, renders the mesogens more mobile, and makes it possible to increase positional disorder of their centers of mass. Such ordering is very sensitive to temperature, and the nematic phase disappears above the \emph{nematic/isotropic transition temperature} $T_{\sf NI}$. As a result, finite temperature changes across the isotropic-nematic transition temperature $T_{\sf NI}$ result into substantial strain of the order of several tens percent, which adds up to the shape change associated to solvent adsorption and desorption, as illustrated in Fig. \ref{Fig_1}.


\medskip
Consider a specimen that, in its preparation state, is in chemical and thermal equilibrium with a solvent at temperature $T_0$ and occupies a region $\Omega$ of the physical space $\mathbb E$. When the specimen is dried and set in thermal equilibrium with a reservoir at temperature $T$, it undergoes a shape change. It was proposed in \cite{Teresi2009} that the combined effect of solvent desorption and temperature change be  described through the theory of elasticity with distortions \cite{NT2007}. According to this theory, the deformation map $\bm f:\Omega\to\mathbb E$ that describes the shape change of the specimen  is a minimizer of the functional
\begin{equation}\label{eq:86}
\mathcal E_0(\bm f):=\int_\Omega \widehat\sigma_0(\bm x,\nabla\bm f^{\sf T}\nabla\bm f){\rm d}V,\qquad \widehat\sigma_0(\bm x,\mathbf C)=\widehat\sigma(\mathbf C\widehat{\mathbf C}_0^{-1}(\bm x,\vartheta)),
\end{equation}
where $\widehat\sigma(\mathbf C)$ is the prototypical strain energy \eqref{eq:104} and where the distortional Cauchy--Green strain $\widehat{\mathbf C}_0(\bm x,\vartheta)$ is a symmetric, positive tensor which depends on both position and renormalized temperature $\vartheta=T/T_{\sf NI}$ (supposed to be spatially uniform), and takes into account both swelling and temperature change.

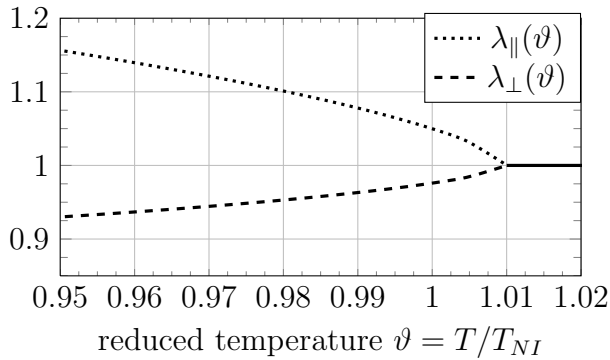
\begin{figure}[h]\centering
\begin{tikzpicture}[scale=1,baseline]
	\begin{axis}	[yscale=0.6,xscale=1,grid=major,minor tick num=3,
	             xmin=0.95,xmax=1.02,
	             ymin=0.85,ymax=1.2,
	             xlabel={reduced temperature $\vth=T/T_{NI}$ },
	             legend style={ at={(1,1.41)}, anchor=east}]
    \addplot[black,dotted,very thick,domain=0.8:1.01,smooth,samples=40] {sqrt(1+2.2*(1.01-x)^(2/3))};
    \addlegendentry{$\lambda_\|(\vartheta)$}
    \addplot[black,dashed,very thick,domain=0.8:1.01,smooth,samples=40]   {1/(1+3.3*2/3*(1.01-x)^(2/3))^(1/4)};
    \addlegendentry{$\lambda_\perp(\vartheta)$}
    \addplot[black,very thick,domain=1.01:1.1,samples=20] {1};
    \end{axis}
 \end{tikzpicture}
 \caption{\label{Fig_23} Plot of $\lamp(\vth)$ and $\lamo(\vth)$ according to (\ref{data_fit}) and \eqref{eq:67}, with $a=2/3$ and $\beta=4.94$ (values taken from \cite{TeresV2013SM}).}
\end{figure}

The expression of the distortional strain is obtained through a multiplicative composition of the separate effects of swelling and temperature change, which result from two experimental evidences. First, when the specimen is dried, it undergoes a relative volume change $0<\nu_{\sf dry}<1$ with respect to the preparation state, accompanied by the distortional strain (here we follow the notation of \cite{TeresV2013SM}):
\begin{equation}
  \label{eq:128}
    \widehat{\mathbf C}_{0,\sf vol}(\bm x)=\alpha_\parallel^2\mathbf N(\bm x)+\alpha_\perp^2(\mathbf I-\mathbf N(\bm x)),
\end{equation}
where $\mathbf N(\bm x)$ is the value attained at $\bm x$ by the nematic tensorial order parameter introduced in \eqref{eq:88} and the stretches $\alpha_\parallel$ and $\alpha_\perp$ are constants satisfying
\begin{equation}
\alpha_\|\alpha_\perp^2 = \nu_{\sf dry}.
\end{equation}
Second, when the specimen's temperature $T$ drops below the isotropic/nematic transition temperature $T_{\sf NI}$, the specimen undergoes a temperature-dependent volume-preserving distortion
\begin{equation}
  \label{eq:124}\begin{split}
   \widehat{\mathbf C}_{0,\sf th}(\bm x,\vartheta)=\lambda_\parallel^2(\vartheta)\mathbf N(\bm x)+\lambda_\perp^2(\vartheta)(\mathbf I-\mathbf N(\bm x)).
\end{split}
\end{equation}
Since thermally- and swelling-induced distortional strains share the same eigenvectors, they commute and determine an overall distortional strain with respect to the preparation state (whose renormalized temperature is $\vartheta_0$)
\begin{equation}
  \label{eq:127}
  \widehat{\mathbf C}_0(\bm x,\vartheta)=\widehat{\mathbf C}_{0,\sf th}(\bm x,\vartheta)\widehat{\mathbf C}^{-1}_{0,\sf th}(\bm x,\vartheta_0)\widehat{\mathbf C}_{0,\sf vol}(\bm x)=\Lambda_\parallel^2(\vartheta)\mathbf N(\bm x)+\Lambda_\perp^2(\vartheta)(\mathbf I-\mathbf N(\bm x)),
\end{equation}
where 
\begin{equation}\label{resultant_stretches}
\Lambda_\|(\vth) = \frac{\lambda_\|(\vth)}{\lambda_\perp(\vartheta_0)}\,\alpha_\|
\quad\text{and}\quad
\Lambda_\perp(\vth) = \frac{\lambda_\perp(\vth)}{\lambda_\perp(\vartheta_0)}\,\alpha_\perp
\end{equation}
are the principal values of the distortion tensor $\mathbf U_0=\sqrt{\mathbf C}_0$. As reported in \cite{TeresV2013SM}, experimental evidence suggests the following expression for $\lambda_\parallel(\vartheta)$:
\begin{equation}\label{data_fit}
\lambda_\parallel(\vartheta)=\left\{
\begin{array}{ll}  \big[ 1+\beta(1.01-\vth)^a \big]^{1/2},&\vth<1, \\[3mm]
                                       1\,,                 &\vth\ge 1\,, 
\end{array}\right.
\end{equation}
where $a$ and $\beta$ are dimensionless parameters and $\vartheta_0=T_0/T_{\sf NI}$ the ratio between the temperature at the preparation state. Once the expression for $\lambda_\parallel(\vartheta)$ has been chosen, the positive function $\lambda_\perp(\vartheta)$ is determined by the condition
\begin{equation}\label{eq:67}
\lambda_\|(\vth)\,\lambda_\perp^2(\vth) = 1,
\end{equation}
which guarantees that the thermal distortional strain $\mathbf C_{0,\sf th}(\vartheta)$ is isochoric. Figure \ref{Fig_23} show the typical temperature dependence of the stretches.

\section{Thin strips of nematic elastomers}\label{sec:3}
In what follows we identify the physical space with $\mathbb R^3$ by fixing a Cartesian reference $(O;x_1,x_2,x_3)$ and we denote by $\{\bm e_i:i=1,2,3$\} the  corresponding orthonormal basis. We consider a specimens that, in its preparation state, occupies a strip--shaped region 
\begin{equation}\label{eq:96}
\Omega=\omega\times(-h/2,+h/2),\qquad 
\end{equation}
of height $h$ (in the $x_3$--direction) and mid-surface $\omega=(-w/2,+w/2)\times (0,L)$, a rectangle of length $L$ and width $w$, respectively, along the $x_1$ and $x_2$ directions. 

If the orientation of the nematic tensor is uniform within the strip, the distortional strain is spatially constant at chemical and thermal equilibrium. In this case, the homogeneous deformation $\bm f_0(\bm x)=\mathbf U_0\bm x$ with $\mathbf U_0=\sqrt{\mathbf C_0}=\Lambda_\parallel(\vartheta)\mathbf N+\Lambda_\perp(\vartheta)(\mathbf I-\mathbf N)$ annihilates the strain energy functional \eqref{eq:86}, and it is not difficult to show that any other smooth minimizer can be obtained from $\bm f_0$ through a surperposed rigid motion. On the other hand, if the nematic director field $\bm d$ depends non-trivially on position, one expect that the response of the specimen to swelling or temperature changes is non-trivial as well. Note, however, that if the renormalized temperature attains the value $\vartheta_{\sf flat}$ that solves the equation
\begin{equation}\label{eq:65}
\Lambda_\|(\vth_{\sf flat})=\Lambda_\perp(\vth_{\sf flat})=:\Lambda^2_{\sf flat},
\end{equation}
then the distortion tensor is spherical and spatially constant, and the energy-minimizing deformation (unique up to a rigid displacement) is the dilation with scaling factor $\Lambda_{\sf flat}$, a deformation that keep the specimen flat, although it alters its dimensions.

In the experiments described in \cite{Sawa2010}, thin strips of nematic elastomers were fabricated with a \emph{splay--bend pattern}, characterized by the nematic orientation being parallel to the plane spanned by the coordinate axes $x_1$ and $x_2$, and changing continuously with respect to the variable $x_3$ from planar alignment (i.e., along the coordinate axis $x_1$) to vertical alignment with respect to the mid surface $\omega$  (i.e., along the coordinate axis $x_3$). The result of such patterning was temperature--dependent bending deformation. Numerical minimization of the energy functional \eqref{eq:86} confirmed the accuracy of the modeling approach outlined in the previous section. 

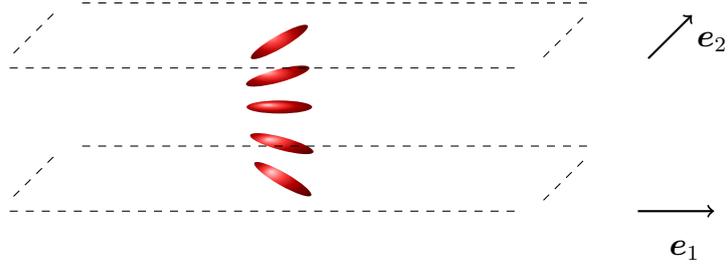
\begin{figure}
\begin{center}
\begin{minipage}{0.8\linewidth}
\begin{center}
\begin{tikzpicture}[scale=1.725]
   \shade[ball color=red, rotate=30] (0.,0.) ellipse (0.25 and 0.05);
   \shade[ball color=red, rotate=15] (-0.08,-0.25) ellipse (0.25 and 0.05);
   \shade[ball color=red, rotate=0] (0,-0.5) ellipse (0.25 and 0.05);
   \shade[ball color=red, rotate=-15] (0.22,-0.75) ellipse (0.25 and 0.05);
   \shade[ball color=red, rotate=-30] (0.55,-0.9) ellipse (0.25 and 0.05);
   
\node (v1) at (-2.14,-1.3) {};
\node (v2) at (1.93,-1.3) {};
\node (v3) at (2.43,-0.8) {};
\node (v4) at (-1.64,-0.8) {};

\node (w1) at (-2.14,-0.2) {};
\node (w2) at (1.93,-0.2) {};
\node (w3) at (2.43,0.3) {};
\node (w4) at (-1.64,0.3) {};
\draw [dashed] (v1) -- (v2) -- (v3) -- (v4) -- (v1);
\draw [dashed] (w1) -- (w2) -- (w3) -- (w4) -- (w1);
\node (v5) at (2.65,-1.3) {};
\node (v6) at (3.4,-1.3) {};
\draw  [thick,->] (v5) edge (v6);
\node at (3.09,-1.6) {\mbox{$\bm e_1$}};
\node (v7) at (2.73,-0.21) {};
\node (v8) at (3.23,0.29) {};
\draw  [thick,->] (v7) edge (v8);
\node at (3.3,0) {$\bm e_2$};
\end{tikzpicture}
\end{center}
\caption{\label{fig1} In the twisted pattern, director is parallel to the plane spanned by $\bm e_1$ and $\bm e_2$. The angle between the nematic director  and and the angle with the axis $\bm e_1$ rotates continuously from $45^\circ$ to $-45^\circ$ from the top to the bottom face of the strip.}
\end{minipage}
\end{center}
\end{figure}

\begin{figure}[!hb]
\begin{center}
    \subfloat[flat shape\label{subfig-1:dummy}]{%
      \includegraphics[width=0.3\textwidth]{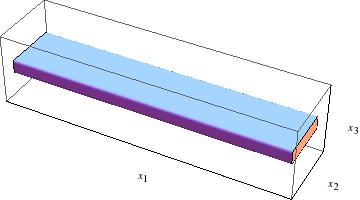}
    }
    \hfill
    \subfloat[helicoid\label{subfig-2:dummy}]{%
      \includegraphics[width=0.3\textwidth]{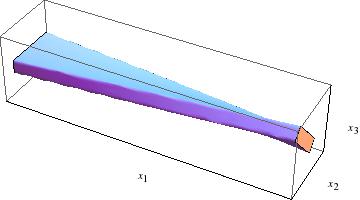}
    }
 \hfill
    \subfloat[spiral\label{subfig-2:dummy}]{%
      \includegraphics[width=0.3\textwidth]{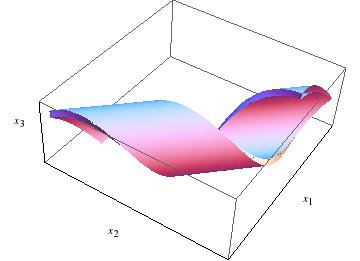}
    }
    \caption{Equilibrium shapes of a thin strip of a nematic elastomer.}
    \label{fig:dummy}
\end{center}
  \end{figure}

In this paper we inspect more closely a different design of the nematic field, referred to as \emph{twisted pattern}, and devised in \cite{sawa2011shape}. In the twisted pattern the nematic orientation is parallel to the mid surface but undergoes an overall rotation of $\pi/2$ about the axis $x_3$, as illustrated in Fig. \ref{fig1}.
The orientation of the nematic director is: 
\begin{equation}\label{eq:90}
 \bm d(\bm x)=\cos(\varphi(x_3))\be_1+\sin(\varphi(x_3))\be_2,\qquad \text{where}\qquad\varphi(x_3)=\frac \pi 4 \frac {x_3}{h/2}.
 \end{equation}
The interest in the twisted pattern stems from the fact that a smooth variation of the ambient temperature $T$ is not accompanied by an equally smooth variation of shape (which is instead the case for the splay--bend pattern). In fact, there exist two values $\vartheta^-_{\sf cr}<\vartheta_{\sf flat}$ and  $\vartheta^+_{\sf cr}>\vartheta_{\sf flat}$ of the (renormalized) temperature such that the specimen switches between helicoidal shape if the renormalized temperature is in the interval $(\vartheta^-_{\sf cr},\vartheta^+_{\sf cr})$ and a spiral shape otherwise, as shown in Fig. \ref{fig:dummy}.

A reduced model based on the F\"oppl--von K\'arm\'an equations with distortions, which exhibits accurate predictions, has been proposed in \cite{sawa2011shape}. The numerical survey carried out in  \cite{TeresV2013SM} confirms such behavior. More recently, a rigorous procedure based on variational convergence \cite{AgostD2015Rigorous} has been used to derive a flexural model. Further dimensional reduction has been performed on this model, taking the cue from \cite{FreddHMP2016JoE,FreddHMP2016apa}, to derive a reduced 1D model \cite{AgostDK2016Shape}. The model we use in this paper, derived in the appendix, features a Koiter--type energy. In the next section we use this model to capture the transition from helicoid to spiral shape.

\section{Helical and spiral configurations}\label{sec:4}
For $\Omega$ the domain defined in \eqref{eq:96}, and for $\bm f$ the deformation map that minimizes the energy \eqref{eq:86}, we focus our attention to the restriction of the deformation map to the mid-surface of the strip:
\[
\bm r(x_1,x_2)=\bm f(x_1,x_2,0),\qquad (x_1,x_2)\in\omega.
\]
Let
\[
{\bm n}=|{\bm r}_{,1}\times {\bm r}_{,2}|^{-1}{\bm r}_{,1}\times {\bm r}_{,2}
\]
denote the unit normal vector field to the parametric surface described by $\bm r(x_1,x_2)$. Starting from the three-dimensional  strain energy \eqref{eq:86} which depends on the entire deformation $\bm f(\cdot)$, we derive in the Appendix the following Koiter--type two-dimensional energy:
\begin{equation}\label{eq:62}
\begin{aligned}
\mathcal E_{\sf 2D}(\bm r)=&\frac h 2 \int_\omega \overline{\mathbb C}_{\alpha\beta\gamma\delta}\left(\frac{a_{\alpha\beta}(\bm r)-a_{\alpha\beta}^0}2\right)\left(\frac{a_{\gamma\delta}(\bm r)-a_{\gamma\delta}^0}2\right){\rm d}A\\
&+\frac {h^3}{24}\int_\omega \overline{\mathbb C}_{\alpha\beta\gamma\delta}(b_{\alpha\beta}(\bm r)-b_{\alpha\beta}^0)(b_{\gamma\delta}(\bm r)-b_{\gamma\delta}^0){\rm d}A,
\end{aligned}
\end{equation}
which depends solely on $\bm r(\cdot)$. Here
\begin{equation}
  \label{eq:118}
  \overline{\mathbb C}_{\alpha\beta\gamma\delta}=2G \delta_{\alpha\gamma}\delta_{\beta\delta}+\overline{k}\delta_{\alpha\beta}\delta_{\gamma\delta},\qquad \overline{k}=\frac{2G}{2G+{k}}{k},
\end{equation}
is the plane elasticity tensor and 
\[
a_{\alpha\beta}(\bm r)=\bm r_{,\alpha}\cdot\bm r_{,\beta},\qquad \text{and}\qquad b_{\alpha\beta}(\bm r)=-\bm n_{,\alpha}\cdot\bm r_{,\beta},
\]
are, respectively, the first and the second fundamental forms of the parametric surface described by $\bm r(\cdot)$. The stretching distortions $a^0_{\alpha\beta}$ and the flexural distortions $b^0_{\alpha\beta}$, which define target metric and the target curvature tensors \cite{dervaux2009morphogenesis,Sharon2009}, have the form
\begin{equation}\label{eq:97}
  a_{\alpha\beta}^0=
\begin{pmatrix}
\Lambda_1^2 & 0\\
0 & \Lambda_2^2
\end{pmatrix},\qquad b_{\alpha\beta}^0=
\begin{pmatrix}
0 & \kappa\\
\kappa & 0\\
\end{pmatrix},
\end{equation}
with the stretches $\Lambda_1$, $\Lambda_2$ and the curvature $\kappa$ being given by the following formulas (here we omit the explicit dependence on temperature): 
\begin{equation}
\Lambda_1^2=\big(\frac{\pi-2}{2\pi}\big)\Lambda_\perp^2+\big(\frac{\pi+2}{2\pi}\big)\Lambda_\parallel^2,\qquad 
\Lambda_2^2=\big(\frac{\pi+2}{2\pi}\big)\Lambda_\perp^2+\big(\frac{\pi-2}{2\pi}\big)\Lambda_\parallel^2,
\end{equation}
and
\begin{equation}\label{eq:64}
\kappa=\frac 6 {\pi^2 h}(\Lambda_\perp^2-\Lambda_\parallel^2).
\end{equation}
The energy density appearing, respectively, in the first and in the second integral on the right--hand side of \eqref{eq:62}, namely,
\[
\pi_{\sf stretch}=\frac h 8 \overline{\mathbb C}_{\alpha\beta\gamma\delta}\left({a_{\alpha\beta}-a_{\alpha\beta}^0}\right)\left({a_{\gamma\delta}-a_{\gamma\delta}^0}\right)
\]
and
\[
\pi_{\sf bend}=\frac{h^3} {24} \overline{\mathbb C}_{\alpha\beta\gamma\delta}\left({b_{\alpha\beta}-a_{\alpha\beta}^0}\right)\left({b_{\gamma\delta}-a_{\gamma\delta}^0}\right)
\]
represent, respectively, the stretching and the bending contribution. Our approach to modeling the spiral--to--helical transition consists in computing the values attained by the functional $\mathcal E_{\sf 2D}$ on two families of deformations, which we refer to as \emph{helicoid configurations} and \emph{spiral configurations}. We shall then determine the transition point by comparing the minima of the energy between the two configurations. From our analysis it will turn out that the transition from helicoid to spiral is ruled by the competition between bending and stretching energy. 

\subsection{Helicoid configurations}
An helicoid configuration with twist $\chi$ is characterized by the typical cross section of the strip undergoing a rotation about the axis $x_3$ by an angle proportional, though $\chi$, to the distance of the cross section from the origin. The corresponding deformation map is 
\begin{equation}
  \label{eq:66}
  \br(\bx)=x_1\be_1+x_2\bR(\chi x_1)\be_2,
\end{equation}
where $\bR(\vartheta)=\be_1\otimes\be_1+(\cos(\theta)\be_2+\sin(\theta)\be_3)\otimes\be_2+(\cos(\theta)\be_3-\sin(\theta)\be_2)\otimes\be_3$ is the counterclockwise rotation tensor of angle $\theta$ about $\be_3$. The components of the first and second fundamental form are given by, respectively,
\begin{equation}
  \label{eq:69}
  \begin{aligned}
    a_{11}(\bx)=1+x_2^2\chi^2,\qquad a_{22}=1,\qquad a_{12}=a_{21}=0,
  \end{aligned}
\end{equation}
and by
\begin{equation}
  \label{eq:70}
  b_{12}(\bx)=b_{21}(\bx)=\frac{\chi}{\sqrt{1+x_2^2\chi^2}},\qquad b_{11}=b_{22}=0.
\end{equation}
Neglecting the contribution of in--plane distortions in the stretching energy we find
\begin{equation}
  \label{eq:81}
\begin{aligned}
  \pi_{\sf stretch,helix}(\chi)&=\frac{h}{2} \frac{G(G+{k}) }{2 G+{k}}x_2^4 \chi ^4.
\end{aligned}
\end{equation}
The bending energy is
\begin{equation}
  \label{eq:89}
\begin{aligned}
  \pi_{\sf bend,helix}(\chi)&=\frac{h^3 G}{6}   \left(\frac{\chi }{\sqrt{1+x_2^2 \chi ^2}}-\kappa\right)^2.
\end{aligned}
\end{equation}
Expanding the bending energy density with respect to $x_2$ and $\chi$, up to the second order, we obtain
\begin{equation}
  \label{eq:87}
   \pi_{\sf bend,helix}(\chi)\simeq \frac{ h^3 G}{6}  (\chi-\kappa)^2
\end{equation}
We see that bending energy scales as $(\chi-\kappa)^2$, whereas stretching energy scales as $\chi^4$. For small values of $\kappa$ it is energetically convenient to minimize the bending energy by taking $\chi=\kappa$. Then, the total energy per unit length is obtained by integrating $\pi_{\sf stretch,helix}(\kappa)$ over the width of the strip. The result is
\begin{equation}
  \label{eq:95}
  W_{\sf helix}(\kappa)=\frac{w^5 h}{160}\frac{ G  (G+{k} )}{2 G+{k} }\kappa^4.
\end{equation}

\subsection{Spiral configurations}
In order to describe spiral configurations, we introduce a new referential coordinate system $(y_1,y_2)$ related to $(x_1,x_2)$ by 
\begin{equation}
  \label{eq:71}
\begin{aligned}
  &y_1(\bx)=\phantom{-}\cos(\psi)x_1+\sin(\psi)x_2,\\
  &y_2(\bx)={-}\sin(\psi)x_1+\cos(\psi)x_2.,
\end{aligned}
\end{equation}
where $\psi$ is an angle to be selected later. Next, for
\begin{equation}
  \label{eq:72}
  {\bf n}(\bx)=\cos(y_1(\bx)/R)\ba_1+\sin(y_1(\bx)/R)\be_3,
\end{equation}
we define
\begin{equation}\
  \label{eq:73}
  \br(\bx)=y_2(\bx)\ba_2+R{\bf n}(\bx).
\end{equation}
It is helpful to think of this configuration as that obtained by rolling the strip over a cylinder of radius $R$, whose axis forms an angle $\psi$ with the axis of the strip. Since ${\bf n}$ is a unit vector satisfying 
\begin{equation}
  \label{eq:76}
  {\bf n}\cdot\ba_2=0,
\end{equation}
we see from \eqref{eq:73} that ${\bf n}\cdot\br_{,\alpha}=y_{2,\alpha}{\bf n}\cdot\ba_2+R{\bf n}\cdot{\bf n}_{,\alpha}=0$. Thus, ${\bf n}$ coincides with the normal to the strip in the deformed configuration. Moreover, as can be easily checked, the deformation map defined by \eqref{eq:73} is an isometry, since the components of the first fundamental form are $a_{11}=a_{22}=1$ and $a_{12}=a_{21}=0$. As a consequence, it is only the bending energy that depends on the variables $\psi$ and $R$.
As to the second fundamental form, we have, by a simple calculation,
\begin{equation}
  \label{eq:74}
  \begin{aligned}
    &{\bf n}_{,1}=-\frac 1 R\cos(\psi)\ba_2\times{\bf n},\\
    &{\bf n}_{,2}=-\frac 1 R\sin(\psi)\ba_2\times{\bf n}.
\end{aligned}
\end{equation}
We then find $b_{11}\stackrel{\rm def}=-{\bf n}_{,1}\cdot\br_{,1}\stackrel{\eqref{eq:73}}=-{\bf n}_{,1}\cdot(y_{2,1}\ba_2+R{\bf n}_{,1})\stackrel{\eqref{eq:76}}=-R{\bf n}_{,1}\cdot{\bf n}_{,1}$, whence, by the first of \eqref{eq:74},
\begin{equation}
  \label{eq:75}
 b_{11}=-\frac 1 R \cos^2(\psi).
\end{equation}
In a similar fashion, we find that the remaining components of the second fundamental form are:
\begin{equation}
  \label{eq:79}
  b_{22}=-\frac 1 R\sin^2(\psi)\qquad{\rm and}\qquad   b_{21}=b_{12}=-\frac 1 R\cos(\psi)\sin(\psi).
\end{equation}
On account of \eqref{eq:79} we have
\begin{equation}
  \label{eq:82}
b_{\alpha\beta}-b^0_{\alpha\beta}=-\frac 1 R \begin{pmatrix}
\cos^2(\psi) & \cos(\psi)\sin(\psi)+R \kappa\\[0.5em]
 \cos(\psi)\sin(\psi)+R \kappa& \sin^2(\psi)
\end{pmatrix}.
\end{equation}
It follows from \eqref{eq:82} that the energy density per unit surface is constant, and is given by 
\begin{equation}
  \label{eq:80}
\begin{aligned}
\pi_{\sf bend,spiral}
&=\frac {h^3}{24}\overline{\mathbb C}_{\alpha\beta\gamma\delta}(b_{\alpha\beta}-b_{\alpha\beta}^0)(b_{\gamma\delta}-b_{\gamma\delta}^0)\\
&=\frac{h^3}{24}\Big(2G (b-b^0)_{\alpha\beta} (b-b^0)_{\alpha\beta}+\frac{2G{k}}{2G+{k}} (b-b^0)_{\alpha\alpha} (b-b^0)_{\beta\beta}\Big)\\
&=\frac 1 {R^2}\frac{h^3}{24}\Big(2G \Big(\cos^4(\psi)+\sin^4(\psi)+2(\cos(\psi)\sin(\psi)+R \kappa)^2\Big)\\
&\qquad\qquad\qquad\qquad\qquad\qquad\qquad\qquad+\frac{2G{k}}{2G+{k}} \big(\cos^2(\psi)+\sin^2(\psi)\big)^2\Big)\\
&=\frac 1 {R^2}\frac{h^3}{24}\Big(2G \Big(\cos^4(\psi)+\sin^4(\psi)+2\cos^2(\psi)\sin^2(\psi)+2(R \kappa)^2+4\cos(\psi)\sin(\psi)R \kappa\Big)\\
&\qquad\qquad\qquad\qquad\qquad\qquad\qquad\qquad\qquad\qquad\qquad\qquad +\frac{2G{k}}{2G+{k}}  \big(\cos^2(\psi)+\sin^2(\psi)\big)^2\Big)\\
&=\frac 1 {R^2}\frac{h^3}{24}\Big(2G \Big((\cos^2(\psi)+\sin^2(\psi))^2+2(R \kappa)^2+4\cos(\psi)\sin(\psi)R \kappa\Big)
\\
&\qquad\qquad\qquad\qquad\qquad\qquad\qquad\qquad\qquad+\frac{2G{k}}{2G+{k}}   \big(\cos^2(\psi)+\sin^2(\psi)\big)^2\Big)\\
&=\frac 1 {R^2}\frac{h^3}{24}\Big(4G\frac{G+{k}}{2G+{k}}+4G R \kappa \big(R \kappa+2\cos(\psi)\sin(\psi)\big)\Big)\\
&=\frac 1 {R^2}\frac{h^3}{24}\Big(4G\frac{G+{k}}{2G+{k}}+4G R \kappa\big(R \kappa+\sin(2\psi)\big)\Big)
\end{aligned}
\end{equation}
The bending energy $\pi_{\sf b}$ is minimized by the following choice of $\psi$ and $R$ find
\begin{equation}
  \label{eq:83}
  \psi=\pm \pi/4,\qquad R=\mp \frac 1 {\kappa}.
\end{equation}
For both choices, the minimum is:
\begin{equation}
  \label{eq:85}
  \begin{aligned}
    \overline\pi_{\sf bend,spiral}(\kappa)&=\frac{h^3}{24}\frac{4G(G+{k})}{2G+{k}}\kappa^2.
  \end{aligned}
\end{equation}
Hence, the energy per unit length of the strip is obtained by integrating $\overline\pi_{\sf bend,spiral}(\kappa)$ over the width of the strip:
\begin{equation}\label{eq:59}
W_{\sf spiral}(\kappa)=w\frac{h^3}{24}\frac{4G(G+{k})}{2G+{k}}\kappa^2.
\end{equation}
\subsection{Switching temperatures}
Formulas \eqref{eq:95} and \eqref{eq:59} are all we need to capture the dependence of the switching temperatures of the geometry of the strip. Indeed, these formulas tell us that if
\[
|\kappa|<\kappa_{\sf switch}:=\sqrt{\frac{80}3}\frac h{w^2},
\]
then the energy of the helicoid configuration, being proportional to $\kappa^4$, is smaller than the energy of the spiral configuration, which is proportional to $\kappa^2$. For $|\kappa|>\kappa_{\sf switch}$, on the other hand, the spiral configuration appears to be energetically more convenient. Thus, we may identify the constant $\kappa_{\sf switch}$ with the (absolute) value of the spontaneous curvature for which the switching between the helical and spiral configuration takes place.

\begin{figure}[h]
\begin{center}
\begin{minipage}{0.8\linewidth}
\begin{center}
\includegraphics{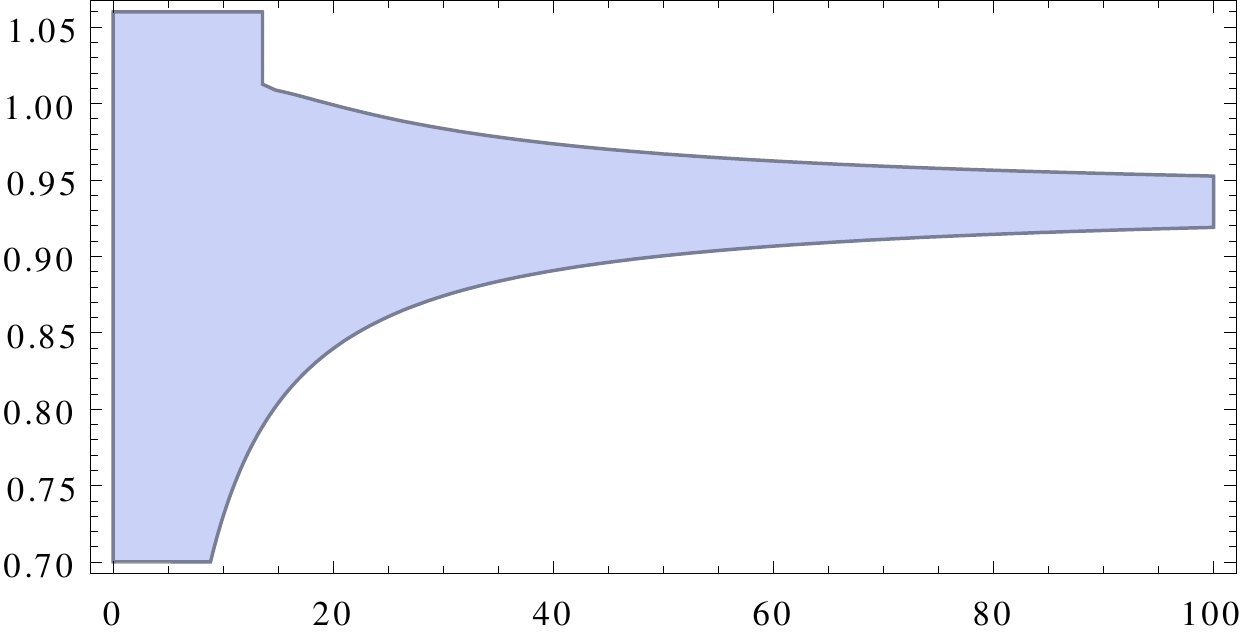}
\end{center}
\caption{\label{fig:5} Phase diagram for the helicoid--spiral transition. We use the ratio $w/h$ as the abscissae, and the renormalized temperature $T/T_{\sf NI}$ as the ordinates. The filled region corresponds to the helical configuration.}
\end{minipage}
\end{center}
\end{figure}

In order to determine the switching temperatures, we restore the dependence of $\kappa$ on $\vartheta$ suppressed in \eqref{eq:64}:
\[
\kappa=\frac 6 {\pi^2 h}(\Lambda_\perp^2(\vartheta)-\Lambda_\parallel^2(\vartheta))=:\widehat\kappa(\vartheta),
\]
and we argue that the helicoid configuration is energetically more convenient than the spiral configuration whenever the renormalized temperature $\vartheta$ satisfies $\widehat\kappa(\vartheta)<\kappa_{\sf switch}$, that is,
\begin{equation}\label{eq:68}
\Lambda_\perp^2(\vartheta)-\Lambda_\parallel^2(\vartheta)<\pi^2 \sqrt{\frac{20}{27}}\left(\frac h w\right)^2.
\end{equation}
That the above inequality is satisfied for $\vartheta=\vartheta_{\sf flat}$, this follows from the fact that 
\[
\widehat\kappa(\vartheta_{\sf flat})=0
\]
(\emph{cf.} \eqref{eq:65}). Since the function $\widehat\kappa(\cdot)$ is continuous, is is possible to identify a maximal interval $(\vartheta_{\sf switch}^-,\vartheta_{\sf switch}^+)$ containing $\vartheta_{\sf flat}$ where the helicoid configuration is energetically cheaper than the spiral configuration. This fact is in accordance with the experimental results reported in \cite{sawa2011shape} and with the numerical calculations carried out in \cite{TeresV2013SM}. Another fact in accordance with \cite{sawa2011shape} and \cite{TeresV2013SM} is that the extreme points of the aforementioned interval, which can be determined by equating the two sides of \eqref{eq:68}, depend on the dimensions of cross section only through the aspect ratio $h/w$. Figure \ref{fig:5} shows a phase diagram obtained using numerical values from Table \ref{Tab1}. This phase diagram is in fairly good agreement with the corresponding one in \cite{sawa2011shape} and shows that the flatter the specimen (\emph{i.e.}, the higher the aspect ratio $w/h$), the smaller is the interval where the helical configuration is stable.  
\begin{table}[h]
\small
\begin{center}
\caption{\ Numerical values used to draw the phase diagram (from \cite{TeresV2013SM})}
\label{Tab1}
\begin{tabular}{lll}
$T_o=313$ K										& preparation temperature \\[2mm]
$T_{NI}=367$ K								    & transition temperature \\[2mm]
$\nu_{\sf dry}$= 0.5 										& dry volume / initial volume \\[2mm]
$\alpha_\|$ = 0.907 							& parallel stretch at dry state\\[2mm]
$\alpha_\perp= \sqrt{v_{\sf dry}/\alpha_\|}$ = 0.743	& orthogonal stretch at dry state\\[2mm]
$a = 2/3\,,\, \beta = 4.94$	 			        & fitting parameter for $\lamp(\vth)$\\[2mm]
$G=6.8\,10^5 \,J/m^3$					        & shear modulus\\[2mm]
$k=1.2\,10^7 \,J/m^3$		      			    & 1st Lam\'e parameter\\[2mm]
\end{tabular}
\end{center}
\end{table}

\section{Appendix}
In this section we derive through a formal calculation the functional \eqref{eq:62} and the expressions \eqref{eq:97} for the distortional first and second fundamental form. As a start, we assume that (here $\bm f_{,i}=\frac{\partial\bm f}{\partial x_i}$) 
\[
\bm f_{,3}(x_1,x_2,0)=\bm n(x_1,x_2),\quad (x_1,x_2)\in\omega.
\]
Under this assumption, the Taylor expansion of the planar components $C_{\alpha\beta}$ (free and dummy Greek indices range between 1 and 2)  of the Cauchy--Green tensor with respect to the referential coordinate $x_3$ is
\begin{equation}\label{eq:77}
C_{\alpha\beta}(\bx)=a_{\alpha\beta}(x_1,x_2)-2x_3 b_{\alpha\beta}(x_1,x_2)+o(x_3),
\end{equation}
where 
\[
a_{\alpha\beta}=\bm r_{,\alpha}\cdot\bm r_{,\beta},\qquad \text{and}\qquad b_{\alpha\beta}=-\bm n_{,\alpha}\cdot\bm r_{,\beta}.
\]
The strain--energy mapping $\widehat\sigma(\cdot)$ vanishes at $\mathbf I$ along with its derivative:
\[
D\widehat\sigma(\mathbf C_0)=\mathbf 0,
\]
that is to say, in components, $\frac{\partial{\widehat\sigma}}{\partial C_{ij}}(\mathbf C_0)=0$. Further computation shows that the second derivative of $\widehat\sigma$ at $\mathbf I$ is the fourth-order tensor
\[
D^2\widehat\sigma(\mathbf I)=\mathbb A,
\]
whose components are $A_{ijkl}=\left.\frac {\partial^2\widehat\sigma}{\partial C_{ij}\partial C_{kl}}\right|_{\mathbf C=I}=\frac {G} 2 \delta_{ik}\delta_{jl}+\frac {k} 4 \delta_{ij}\delta_{kl}$. As a result, we obtain 
\begin{equation}\label{eq:92}
\widehat\sigma(\mathbf C\mathbf C_0^{-1})=\widehat\sigma_{\sf Q}(\mathbf C-\mathbf C_0)+o(|\mathbf C-\mathbf I|^2+|\mathbf C_0-\mathbf I|^2), 
\end{equation}
where
\[
\widehat\sigma_{\sf Q}(\mathbf C-\mathbf C_0):=\frac 12 \mathbb A[\mathbf C-\mathbf C_0]\cdot(\mathbf C-\mathbf C_0).
\]
Under the assumption that both the Cauchy--Green strain $\mathbf C=\nabla\bm f^{\sf T}\nabla\bm f$ and the distortion tensor $\mathbf C_0$ depart modestly from the identity $\mathbf I$, we discard the last term on the right--hand side on \eqref{eq:92}. We next augment the partial representation \eqref{eq:77} for the Cauchy--Green strain with the assumption that the second Piola--Kirchhoff stress associated to the quadratic energy $\widehat\sigma_{\sf Q}$:
\[
\bm\Sigma=2D\widehat{\sigma}_{\sf Q}(\mathbf C-\mathbf C_0)= 2\mathbb A(\mathbf C-\mathbf C_0)
\]
satisfies the plane--stress assumption:
\[
\Sigma_{3i}=0.
\]
Such assumption determines the components $C_{3i}$ of the Cauchy--Green strain and allows us to write the quadratic energy $\sigma_{\sf Q}$ as:\footnote{Notice that the energy \eqref{eq:94} depends only on the planer components $C_{\alpha\beta}$ of the Cauchy--Green strain. Moreover, such energy \eqref{eq:94} coincides with the relaxation ($\mathcal V$ is the translation space of $\mathbb E$):
\[
\overline\sigma_{\sf Q}(\mathbf A)=\min_{\mathbf a\in\mathcal V}\frac 12 \mathbb A[\mathbf A+{\rm sym}(\mathbf a\otimes\be_3)]\cdot[\mathbf A+{\rm sym}(\mathbf a\otimes\be_3)]
\]
which appears in rigorous deductions of nonlinear plate models \cite{FriesJM2002CPAM,Schmi2007Plate,AgostD2015Rigorous}.}
\begin{equation}\label{eq:94}
\sigma_{\sf Q}=\frac 12 \overline{\mathbb A}[\mathbf C-\mathbf C_0]\cdot(\mathbf C-\mathbf C_0),
\end{equation}
where $\overline{\mathbb A}_{3kij}=\overline{\mathbb A}_{k3ij}=\overline{\mathbb A}_{ij3k}=\overline{\mathbb A}_{ijk3}=0$ and
\begin{equation}
  \label{eq:118}
  \overline{\mathbb A}_{\alpha\beta\gamma\delta}=\frac {G} 2\delta_{\alpha\gamma}\delta_{\beta\delta}+\frac{\overline{k}}4\delta_{\alpha\beta}\delta_{\gamma\delta},\qquad \overline{k}=\frac{2{G}}{2{G}+{k}}{k}.
\end{equation}
By making use of \eqref{eq:77} we obtain
\[
\int_\Omega \sigma_{\sf Q}{\rm d}V=\int_\omega \sigma_{\sf 2D}{\rm d}A,
\]
where, on setting 
\begin{equation}\label{eq:91}
  g^0_{\alpha\beta}=\mathbf C_0\cdot\be_\alpha\otimes\be_\beta,
\end{equation}
we have
\begin{equation}
  \label{eq:119}
\begin{split}
  \sigma_{\sf 2D}=&\int_{-h/2}^{+h/2}\frac 12 \overline{\mathbb A}[{\mathbf C}-{\mathbf C}_0]\cdot
({\mathbf C}-{\mathbf C}_0){\rm d}x_3
\\
=&\int_{-h/2}^{+h/2}\frac 12 \overline{\mathbb A}_{\alpha\beta\gamma\delta}
(a_{\alpha\beta}-2x_3b_{\alpha\beta}-g^0_{\alpha\beta})(a_{\gamma\delta}-2x_3b_{\gamma\delta}-g^0_{\gamma\delta})
\\
=&
{\int_{-h/2}^{+h/2}\frac 12\overline{\mathbb A}_{\alpha\beta\gamma\delta}
(a_{\alpha\beta}-g^0_{\alpha\beta})
(a_{\gamma\delta}-g^0_{\gamma\delta}){\rm d}x_3}+{\int_{-h/2}^{+h/2} 2 x_3 \overline{\mathbb A}_{\alpha\beta\gamma\delta} b_{\alpha\beta}(g^0_{\gamma\delta}+x_3 b_{\gamma\delta}){\rm d}x_3}
+o(h^3),
\\
=&\frac h 2 \bar{\mathbb A}_{\alpha\beta\gamma\delta}(a_{\alpha\beta}-a_{\alpha\beta}^0)(a_{\gamma\delta}-a_{\gamma\delta}^0)+\frac {h^3}{6}\bar{\mathbb A}_{\alpha\beta\gamma\delta}(b_{\alpha\beta}-b_{\alpha\beta}^0)(b_{\gamma\delta}-b_{\gamma\delta}^0)+C+o(h^3),
\end{split}
\end{equation}
where $C$ is a constant that depends on $g_{\alpha\beta}^0$, but not on the deformation, and
\begin{equation}
  \label{eq:123}
  a^0_{\alpha\beta}:=\langle g^0_{\alpha\beta}\rangle,\qquad\text{and}\qquad b^0_{\alpha\beta}=\frac 6 {h^2}\big\langle x_3 g^0_{\alpha\beta}\big\rangle
\end{equation}
and with chevrons denoting over--the--thickness average: $\langle\varphi\rangle=\frac 1 h \int_{-h/2}^{+h/2}\varphi{\rm d}x_3$. On setting
\[
\overline{\mathbb C}=4\overline{\mathbb A},
\]
and on computing the thickness averages that define $a^0_{\alpha\beta}$ and $b^0_{\alpha\beta}$ we obtain the two--dimensional energy \eqref{eq:62} and the formulas \eqref{eq:97}--\eqref{eq:64}.
\bigskip

\noindent\textbf{Acknowledgments}

\noindent Both authors thank Luciano Teresi for several fruitful discussions. GT thanks Alessandro Lucantonio for pointing out some relevant references.\medskip\null\medskip

\noindent\textbf{Compliance with Ethical Standards}

\noindent \emph{Funding}: The authors are supported by INdAM-GNFM through grant ``Progetto Giovani 2016: Mathematical modelling of bio-hybrid and bio-inspired soft robots''.\smallskip

\noindent \emph{Conflict of Interest}: The authors declare that they have no conflict of interest.

\bibliographystyle{abbrv}
\bibliography{bibliography-gt,Biblio_TNEs}

\end{document}